\documentclass[runningheads]{llncs}
\usepackage{float}
\usepackage{graphicx}
\usepackage{hyperref}
\usepackage{caption}
\usepackage{subcaption}
\usepackage{diagbox}
\usepackage[dvipsnames]{xcolor}
\begin{document}
\title{Risk-dependent centrality in the Brazilian stock market}
\author{Michel Alexandre\inst{1,2} \and
Kauê L. de Moraes\inst{3} \and
Francisco A. Rodrigues\inst{2}}
\authorrunning{M. Alexandre et al.}
\institute{Central Bank of Brazil \and
Institute of Mathematics and Computer Science, University of São Paulo, Brazil \and Department of Economics, University of São Paulo, Brazil}
\maketitle              % typeset the header of the contribution
\begin{abstract}
The purpose of this paper is to calculate the risk-dependent centrality (RDC) of the Brazilian stock market. We computed the RDC for assets traded on the Brazilian stock market between January 2008 to June 2020 at different levels of external risk. We observed that the ranking of assets based on the RDC depends on the external risk. Rankings' volatility is related to crisis events, capturing the recent Brazilian economic-political crisis. Moreover, we have found a negative correlation between the average volatility of assets' ranking based on the RDC and the average daily returns on the stock market. It goes in hand with the hypothesis that the rankings' volatility is higher in periods of crisis.

\end{abstract}
\section{Introduction}
\begin{figure}[H]
    \centering
    \includegraphics[width=\textwidth]{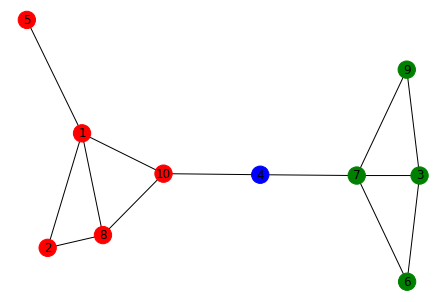}
    \caption{The removal of node 4 will break the network into two disconnected parts. Moreover, all shortest paths from one region to another has to go through node 4.}
    \label{fig:fig1}
\end{figure}

Centrality is one of the key topological features of nodes in complex networks ~\cite{Rodrigues019,Comin2020}. It refers to how important and influential a given node is for the whole network. Central nodes play an important role concerning dynamical processes throughout complex networks, e.g., the spreading of news in social networks or the propagation of shocks in financial networks. Moreover, the removal of a central node is supposed to cause a significant change in the structure of the network; otherwise (for instance, if it is a dead-end node), its removal will have no further consequences. Consider the simple example depicted in Figure \ref{fig:fig1}. The removal of node 4 will break the network into two disconnected components, implying that 4 is a very central node. Notice that this node is not highly connected and, therefore, not only hubs are considered as central.

\paragraph{} There are many centrality measures to characterize the networks structure~\cite{Rodrigues019}. Particularly, three classical ones are (i) the degree (the number of direct neighbors of a given node), (ii) the betweenness centrality (the fraction of shortest paths\footnote[1]{The shortest path between two nodes is the one in which the sum of the weights of the constituent edges is minimized. There are some excellent textbooks the reader unfamiliar with the complex networks concepts may refer to, as \cite{estrada2012} and \cite{barabasi2016}.} going through a given node), and (iii) the closeness centrality (the average of the shortest path from a given node to every other node in the network).\footnote[2]{They are note necessarily correlated, hence the necessity of considering more than one measure of centrality. Coming back to Figure \ref{fig:fig1}, node 4 has a small degree (just two direct neighboring nodes). However, all shortest paths from the region comprised of the red nodes to the region comprised of the green nodes has to go through node 4, which implies it has a high value of betweenness centrality.} The eigenvector centrality (\cite{bonacich1972ec}) is not restricted to shortest paths. It is defined as the main eigenvector of the adjacency matrix representing the network. Other centrality measures include the subgraph centrality (\cite{estrada2005sc}), and the PageRank centrality (\cite{gleich2015}).
\paragraph{}The use of centrality measures in the analysis of economic and financial networks has proved to be quite fruitful. It has shed light on a wide range of important issues. Several studies (e.g., \cite{jaramillo2014}, \cite{kuzubas2014}, \cite{leon2014}, \cite{chanlau2018}, \cite{ghan2018}) have attested the relevance of centrality measures in identifying systemically important nodes in financial networks. D'Errico et al. \cite{derrico2009} studied the shareholding network of the Italian Stock Market. By using some centrality measures, they detected central companies according to two criteria: informational flow and absorption of external shocks (less asset volatility in central companies). Rossi et al. \cite{rossi2018} found a positive correlation between network centrality and portfolio performance in a delegated investment management setting. Temizsoy et al. \cite{temiz2017} show that funding rates in the e-MID market are significantly affected by banks' network centrality in the interbank market. The temporal centrality measure, developed by \cite{zhao2018}, proved to be an efficient portfolio optimization and risk management tool.
\paragraph{}A shortcoming of the existent measures of centrality is that they are based on a static view of the network. Networks are subject to elements that can cause significant changes in relatively short periods of time. For instance, changes in the external risk will alter the centrality ranking of the banks in an interbank network (\cite{bartesaghi2020}). Based on the Susceptible-Infected epidemiological models (\cite{pastor2015}) and on the communicability functions (\cite{estrada2008}), Bartesaghi et al. (\cite{bartesaghi2020}) derived a new centrality measure called \textit{risk-dependent centrality} (RDC). This centrality measure, as well as the ranking of the nodes in the network according to this index, vary with the change in the external risk the whole network is submitted to.
\paragraph{} The purpose of this paper is to calculate the RDC of the Brazilian stock market. Following \cite{bartesaghi2020}, we extracted the minimum spanning tree (MST) from the correlation matrix of daily returns of assets traded at the Brazilian stock market. The period ranges from January 2008 to June 2020. The RDC is computed for each asset at different levels of external risk. We observed that the ranking of assets based on the RDC depends on the external risk: assets which are central at low levels of external risk are less central when this level increases, and vice-versa. Moreover, we have found a negative correlation between the average volatility of assets' ranking based on the RDC and the average daily returns on the stock market. It goes in hand with the hypothesis that the rankings' volatility is higher in periods of crisis.
\paragraph{} Besides this introduction, this paper has three other parts. Section \ref{sec:meth} presents the methodology and the data set. Results are discussed in Section \ref{sec:res}. Some concluding remarks take Section \ref{sec:concl}.
\section{Methodology and data set}
\label{sec:meth}
\subsection{The data set}
\label{subs:data}
\begin{figure}[H]
    \centering
    \includegraphics[width=100mm]{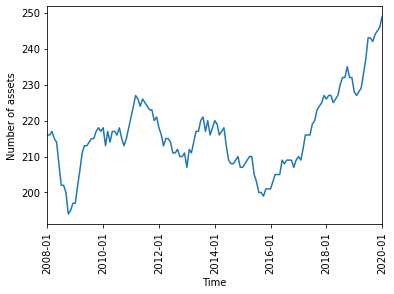}
    \caption{Number of assets used in each time window. The time period in the horizontal axis corresponds to the first month of the time window.}
    \label{fig:nass}
\end{figure}

Our data set is comprised of daily closure prices of the assets traded at the Brazilian stock market (BMF\&BOVESPA) between January 2008 and June 2020. Following Bartesaghi et al. (\cite{bartesaghi2020}), we used moving six-months windows, in which the window is rolled one month forward. The first time window covers all trading days from January $1^{st}$, 2008 to June $30^{th}$, 2008; the second one, from February $1^{st}$, 2008 to July $31^{st}$, 2008; and so on, until the last time window, from January $1^{st}$, 2020 to June $30^{th}$, 2020. Hence, we have obtained a total of 145 time windows.
\paragraph{}In each window, we have considered only those assets whose number of observations is at least 80\% of the number of trading days in the period. This is to assure that the returns correlation can be properly calculated. The number of assets in each time window varies from 196 to 249 and is depicted in Figure \ref{fig:nass}.

\subsection{Methodological issues}
\label{subs:meth}
\begin{figure}[H]
    \centering
    \includegraphics[width=70mm]{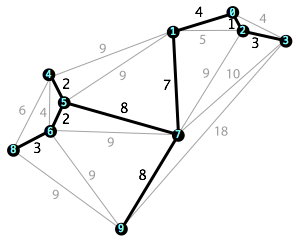}
    \caption{A graph and its MST represented by the edges in bold.}
    \label{fig:mst}
\end{figure}

 Once the data was retrieved, we first calculated the logarithmic returns of the assets:
\begin{equation}
r_i^{t} = {\ln P_i^t} - {\ln P_i^{t-1}} .
\label{eq:ret}
\end{equation}
\paragraph{}
In Equation \ref{eq:ret} above, $P_i^t$ is the closure price of asset \textit{i} at time t. Then we computed the Pearson cross correlation coefficient $\rho_{ij}$ between the returns of each pair of assets. Finally, we transformed the correlation coefficients $\rho_{ij}$ into the distance coefficients $d_{ij}$ according to the following equation (\cite{mantegna1999,Peron011,Peron2012}):
\begin{equation}
d_{ij} = \sqrt{2(1 - \rho_{ij})} .
\label{eq:dist}
\end{equation}
\paragraph{}The higher the correlation between the returns of two different assets, the closer they will be. In fact, $d_{ij}$ reaches its minimal value (0) when $\rho_{ij}$ is 1, and the distance will be maximal (2) when the correlation is minimal (-1). The coefficients $d_{ij}$ form the distance matrix \textit{D}, which constitutes the adjacency matrix of graph $\Gamma$.
\paragraph{}The minimum spanning tree (MST) \textit{T} is extracted from graph $\Gamma$. A spanning tree connects all the nodes of the graph without forming loops. Among all existing spanning trees, that whose sum of weights of the edges is the lowest is MST. Figure \ref{fig:mst} illustrates this point. The weights of the edges of the MST (in bold) amounts to 38. It can be seen that it is not possible to connect all the vertices of the graph through a shortest path of edges and without loops. MST is an approach extensively applied to financial networks (e.g., \cite{gilmore2008}, \cite{kwapien2017}), as it highlights the most relevant piece of a graph, bearing only its essential information.

\begin{figure}[H]
    \centering
    \includegraphics[width=70mm]{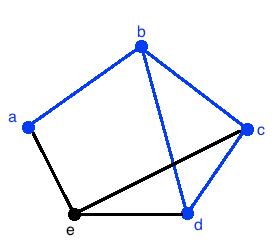}
    \caption{The path \textit{abcdb} is a possible walk starting at \textit{a} and ending at \textit{b} (length=4).}
    \label{fig:walk}
\end{figure}

\paragraph{}The final step is the calculation of the RDC. In the undirected network represented by the MST, the RDC of each node \textit{i} is calculated according to the following equation:\footnote[3]{For more details on the RDC formula, see \cite{bartesaghi2020}.}

\begin{equation}
    R_i = \sum_{k=1}^{\infty} \zeta^{k}\frac{A^{k}}{k!} = e^{\zeta A} ,
\label{eq:rdc}
\end{equation}

where \textit{A} is the adjacency matrix representing the MST and $A^k$ is the number of walks of length k starting at \textit{i} to the power of k.\footnote[4]{In order to solve Equation \ref{eq:rdc}, we rely on the following properties of symmetric matrices (as this is the case of \textit{A}): matrix \textit{A} can be decomposed as $A = Q \Lambda Q^{-1}$; and all \textit{A} eigenvalues are real. It implies the following exponential matrix property: $e^{\zeta A} = Qe^{\zeta \Lambda}Q^{-1}$. For details, see \cite{moler2003}.} The parameter $\zeta$ corresponds to the exogenous level of external risk to which the whole network is submitted. Figure \ref{fig:walk} depicts a simple network composed of five nodes. The path \textit{abcdb} is a possible walk starting at node \textit{a} and ending at node \textit{b}. Assuming all edges are of length 1, this walk is of length 4. The RDC is the sum of two components: the \textit{circulability}, i.e., the weighted sum of all closed walks starting and ending at node \textit{a}, and the \textit{transmissibility}, the weighted sum of all closed walks that start at \textit{a} and end elsewhere.

\section{Results}
\label{sec:res}
\subsection{General results}
\begin{figure}[H]
\centering
    \begin{subfigure}[t]{\textwidth}
        \centering
        \includegraphics[width=100mm]{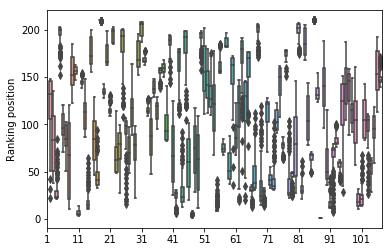}
    \end{subfigure}
    \newline
    \begin{subfigure}[t]{\textwidth}
        \centering
        \includegraphics[width=100mm]{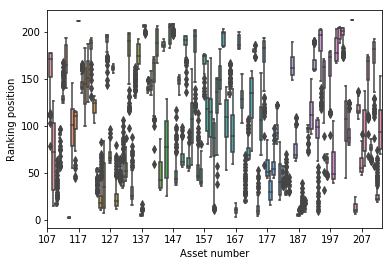}
    \end{subfigure}
    \caption{Distribution of assets' rankings based on the RDC with respect to $\zeta$. Results regard to the time window 2008-01 (January 2008-June 2008). For the sake of better visualization, results are being presented in two subplots.}
    \label{fig:rank}
\end{figure}

For each time window, we followed the steps described in Section \ref{subs:meth}. The ranking position of each asset based on the RDC is computed varying the value of the external risk $\zeta$ in the interval (0,1] with step 0.01.
\paragraph{}The result for the time window 2008-01 (January 2008-June 2008) is depicted in Figure \ref{fig:rank}. In most cases, the ranking position changes significantly according to different values of $\zeta$. Some assets become less central when the external risk increases. For instance, TIM (telecommunications, number 192 in Figure \ref{fig:rank}) moved from position 9 to 107 when $\zeta$ increases from 0.01 to 1. In contrast, other assets climbed in the ranking with the increase in $\zeta$, as \textit{F. Guimarães} (textile, number 79, from $146^{th}$ to $27^{th}$ place) and \textit{Embratel} (telecommunications, number 68, from $119^{th}$ to $15^{th}$ place). Other assets are very central even under high levels of external risk. The airline \textit{Gol} (number 87) always occupies the first place, regardless the value of $\zeta$. Other quite central assets are, for instance, the electric power company \textit{Light} (number 113, worst ranking equal to 3) and the gas distribution company \textit{Comgas} (number 46, worst ranking equal to 5). On the other hand, some assets have a very low level of centrality independently of the external risk. The companies \textit{Vale} (mining, number 203), \textit{Lojas Americanas} (retail, number 116), \textit{B2W Digital} (e-commerce, number 17), and \textit{Gerdau} (metallurgy, number 85) are always among the four least central assets.

\begin{figure}[H]
    \centering
    \includegraphics[width=120mm]{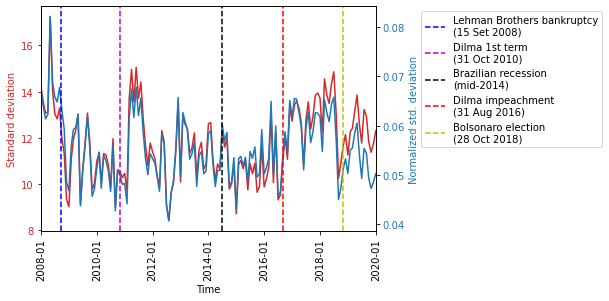}
    \caption{Average standard deviation of assets' rankings based on the RDC. The time period in the horizontal axis corresponds to the first month of the time window.}
    \label{fig:sd}
\end{figure}

\paragraph{}We calculated the average standard deviation of assets' rankings based on the RDC for each time window. As each time windows has a different number of assets, we computed the standard deviation normalized according to the number of assets. This avoids any size effects and allows the comparison between different time windows. Both measures are shown in Figure \ref{fig:sd}. There was a peak of rankings' volatility in the midst of the 2007-2008 financial crisis, whose climax was the Lehman Brothers bankruptcy. The period between mid-2014 and end-2018, during which Brazil experienced a harsh economic-political crisis, is also characterized by a roughly continuous increase in rankings' volatility.   

%\subsection{Financial x non-financial firms}
%\paragraph{}Figure \ref{fig:sd_sec} depicts the average normalized standard deviation of assets' rankings for financial and non-financial firms. It can be seen that standard deviation is more volatile over time for financial firms than for non-financial ones. In Figure \ref{fig:sd_av}, we present the average ranking for financial and non-financial firms. As the average ranking is highly dependent on the number of assets, we also normalize the average according to this quantity. Both types of firms display a very different dynamics: while non-financial firms present a quite stable average ranking centrality, that of financial firms oscillates widely over time. In most of the time windows, the average ranking of non-financial firms is smaller than that of financial firms (i.e., the former are more central than the latter). However, recently, it seems that financial firms became more central. It suggests that the ranking position based on RDC of financial firms is much more impacted by external shocks than that of non-financial firms.

%\begin{figure}[H]
%    \centering
%    \includegraphics[width=100mm]{sector_stdev2.png}
%    \caption{Average normalized standard deviation of assets' rankings based on the RDC for financial and non-financial firms.}
%    \label{fig:sd_sec}
%\end{figure}

%\begin{figure}[H]
%    \centering
%    \includegraphics[width=100mm]{sector_mean2.png}
%    \caption{Average assets' rankings based on the RDC for financial and non-financial firms.}
%    \label{fig:sd_av}
%\end{figure}

\subsection{RDC and crisis}
\paragraph{}According to \cite{bartesaghi2020}, the centrality of nodes is more affected by $\zeta$ in shock periods. Therefore, the average standard deviation of assets' rankings based on the RDC should be higher in periods of crisis. One of the main characteristics of financial crisis is a strong depreciation of assets, leading stock market indexes to plunge. In the time windows subsequent to the 2008 financial crisis, the average daily return of the main Brazilian stock market index (\textit{Ibovespa}) was near -0.4\%.

%\begin{figure}[H]
%    \centering
%    \includegraphics[width=100mm]{retd.png}
%    \caption{\textit{Ibovespa} average daily return (in \%).}
%    \label{fig:retd}
%\end{figure}

\begin{figure}[]
\centering
    \begin{subfigure}[t]{\textwidth}
        \centering
        \includegraphics[width=100mm]{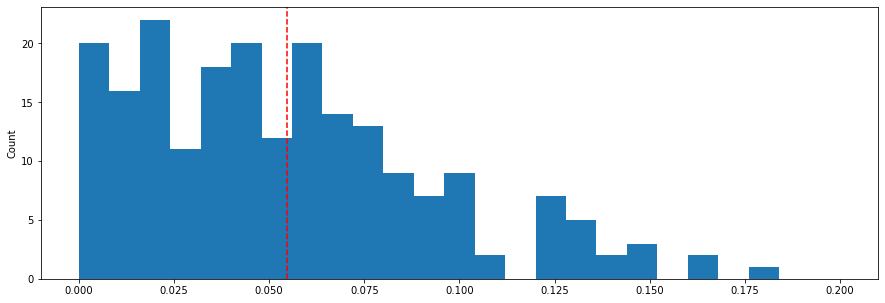}
    \end{subfigure}
    \newline
    \begin{subfigure}[t]{\textwidth}
        \centering
        \includegraphics[width=100mm]{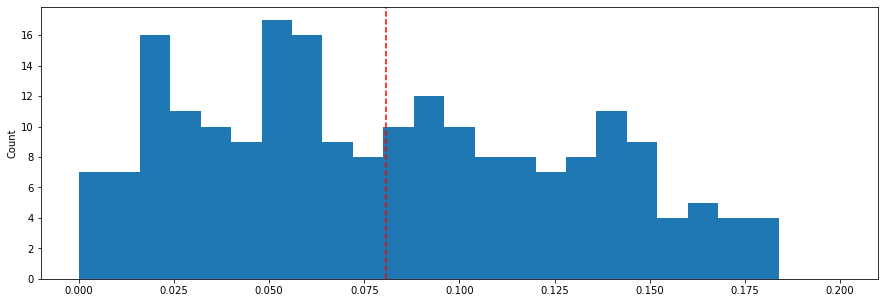}
    \end{subfigure}
    \caption{Distribution of normalized standard deviation of assets' rankings based on the RDC for two time windows: with the highest average \textit{Ibovespa} daily return, 2009-10 (above), and with the lowest one, 2008-05 (bellow). The dashed lines represent the average standard deviation in each time window.}
    \label{fig:hist}
\end{figure}

\paragraph{}We computed the correlation between the average standard deviation of assets' rankings based on the RDC and the \textit{Ibovespa} average daily return. As expected we have found a negative correlation. This correlation is more more significant (both in terms of absolute value and p-value) when the normalized standard deviation is considered (Table \ref{tab:corr}).
\paragraph{}To illustrate this point, we present in Figure \ref{fig:hist} the distribution of the normalized standard deviation of assets' rankings according to the RDC for two time windows. We chose the time window 2009-10, that had the highest average \textit{Ibovespa} daily return (0.47\%), and 2008-05, with the lowest (-0.44\%). As expected, the average standard deviation is smaller in the first case (0.055 against 0.081), with a p-value around $1.8\times10^{-9}$.   

\begin{table}[H]
    \centering
    \begin{tabular}{|l||c|c|}
    \hline
    \makebox{}& \makebox{correlation} & \makebox{p-value} \\
    \hline \hline
    Standard deviation & -0.1762 & 0.0340 \\
    \hline
    Normalized std. dev. & -0.2207 & 0.0076\\
    \hline
    \end{tabular}
    \caption{Correlation between the average standard deviation of assets' rankings based on the RDC and the \textit{Ibovespa} average daily return.}
    \label{tab:corr}
\end{table}

\paragraph{}As a last exercise, we ranked the time windows according to the \textit{Ibovespa} average daily return. We formed two groups: the five time windows with the highest and the lowest average daily return (top 5 and bottom 5, respectively). Finally, we compared the average normalized standard deviation of assets' rankings based on the RDC between time windows of different groups. The results are presented in Table \ref{tab:ttest}. As expected, no time window in the bottom 5 group has an average standard deviation smaller than any time window in the top 5 group. Moreover, in the majority of the cases, the p-value is smaller than 5\%. 

\begin{table}[H]
    \centering
    \bigskip
    \noindent
    \begin{tabular}{|l||*{5}{l|}}\hline
    \backslashbox{Top 5}{Bottom 5}
    &\makebox[4em]{2008-05}&\makebox[4em]{2008-09}&\makebox[4em]{2008-07}
    &\makebox[4em]{2008-06}&\makebox[4em]{2008-08}\\\hline\hline
    2009-10 &\textcolor{red}{0.026}&\textcolor{red}{0.012}&\textcolor{green}{0.010}&\textcolor{red}{0.013}&\textcolor{green}{0.009}\\\hline
    2009-11 &\textcolor{red}{0.036}&\textcolor{red}{0.022}&\textcolor{red}{0.020}&\textcolor{red}{0.023}&\textcolor{red}{0.019}\\\hline
    2009-07 &\textcolor{red}{0.031}&\textcolor{red}{0.017}&\textcolor{red}{0.015}&\textcolor{red}{0.018}&\textcolor{red}{0.014}\\\hline
    2016-11 &\textcolor{red}{0.027}&\textcolor{red}{0.013}&\textcolor{red}{0.011}&\textcolor{red}{0.014}&\textcolor{green}{0.010}\\\hline
    2009-09 &\textcolor{red}{0.020}&0.006&0.004&0.007&0.003\\\hline
    \end{tabular}
    \bigskip
    \caption{The values in the table correspond to the difference between the average standard deviation of assets' rankings based on the RDC of the time window in the bottom 5 group (columns) and that of the time window in the top 5 group (rows). \textcolor{red}{Red}: significant at the 1\% level; \textcolor{green}{Green}: significant at the 5\% level.}
    \label{tab:ttest}
\end{table}

\section{Concluding remarks}
\label{sec:concl}
We computed the RDC for the Brazilian stock market, replicating the results presented in \cite{bartesaghi2020} for the U.S. stocks constituents of the \textit{S\&P} 100 index. In most of the cases, not only the RDC, but also the assets' rankings based in it, change significantly with the level of external risk. Assets lose or gain positions in the centrality ranking when the external risk level varies. Few assets remain at the top or at the tail of the ranking independent of the external risk.
%\paragraph{}We also have shown the results considering two types of assets: those of financial and non-financial firms. Interestingly, we could observe that the average ranking according to the RDC of financial firms is much more volatile over time than that of non-financial firms. It suggests that the ranking position based on the RDC of financial firms is much more dependent on the external risk level than that of non-financial firms. 
\paragraph{}Additionally, we assessed the relationship between crisis and RDC. Bartesaghi et al. (\cite{bartesaghi2020}) pointed that the centrality of nodes is more affected by the external risk in periods of crisis. Comparing the Jan-Jun 2001 time window with the two networks covering the 2007-2008 crisis period (end of 2007 and end of 2008), they show that the volatility of assets' ranking based on the RDC is significantly higher in the first case. We tested this hypothesis through an alternative approach. As crises are characterized by a plunge in the stock market indexes, we investigated the relationship between the volatility of assets' rankings based on the RDC and the stock market index return. We found a negative correlation between the average standard deviation of assets' rankings and the \textit{Ibovespa} index average daily return, mainly when we considered the standard deviation normalized by the number of assets. Moreover, tests of mean difference have shown that time windows in the bottom 5 of average daily return present an average normalized standard deviation higher than that of those in the top 5. These results support the above-mentioned hypothesis.

\bibliographystyle{splncs04}
\bibliography{samplepaper}

\begin{thebibliography}{10}
\providecommand{\url}[1]{\texttt{#1}}
\providecommand{\urlprefix}{URL }
\providecommand{\doi}[1]{https://doi.org/#1}

\bibitem{barabasi2016}
Barabási, A.L.: Network science. Cambridge University Press (2016)

\bibitem{bartesaghi2020}
Bartesaghi, P., Benzi, M., Clemente, G.P., Grassi, R., Estrada, E.:
  Risk-dependent centrality in economic and financial networks. SIAM Journal of
  Financial Mathematics  \textbf{11}(2),  526--565 (2020)

\bibitem{bonacich1972ec}
Bonacich, P.: Factoring and weighting approaches to status scores and clique
  identification. Journal of Mathematical Sociology  \textbf{2}(1),  113--120
  (1972)

\bibitem{chanlau2018}
Chan-Lau, J.A.: Systemic centrality and systemic communities in financial
  networks. Quantitative Finance and Economics  \textbf{2}(2),  468--496 (2018)

\bibitem{Comin2020}
Comin, C.H., Peron, T., Silva, F.N., Amancio, D.R., Rodrigues, F.A., Costa,
  L.d.F.: Complex systems: Features, similarity and connectivity. Physics
  Reports  (2020)

\bibitem{derrico2009}
D’Errico, M., Grassi, R., Stefani, S., Torriero, A.: Shareholding networks
  and centrality: An application to the italian financial market. In: Naimzada,
  A.K., Stefani, S., Torriero, A. (eds.) Networks, Topology and Dynamics:
  Theory and Applications to Economics and Social Systems, pp. 215--228.
  Springer, Berlin (2009)

\bibitem{estrada2008}
Estrada, E., Hatano, N.: Communicability in complex networks. Physical Review E
   \textbf{77}(3),  036111 (2008)

\bibitem{estrada2012}
Estrada, E.: The structure of complex networks: theory and applications. Oxford
  University Press (2012)

\bibitem{estrada2005sc}
Estrada, E., Rodríguez-Velázquez, J.A.: Subgraph centrality in complex
  networks. Physical Review E  \textbf{71}(5),  056103 (2005)

\bibitem{ghan2018}
Ghanbari, R., Jalili, M., Yu, X.: Correlation of cascade failures and
  centrality measures in complex networks. Future Generation Computer Systems
  \textbf{83},  390--400 (2018)

\bibitem{gilmore2008}
Gilmore, C., Lucey, B., Boscia, M.: An ever-closer union? examining the
  evolution of linkages of european equity markets via minimum spanning trees.
  Physica A: Statistical Mechanics and its Applications  \textbf{387}(25),
  6319--6329 (2008)

\bibitem{gleich2015}
Gleich, D.F.: Page{R}ank beyond the web. Siam Review  \textbf{57}(3),  321--363
  (2015)

\bibitem{Peron2012}
Kau{\^e} Dal’Maso~Peron, T., da~Fontoura~Costa, L., Rodrigues, F.A.: The
  structure and resilience of financial market networks. Chaos: An
  Interdisciplinary Journal of Nonlinear Science  \textbf{22}(1),  013117
  (2012)

\bibitem{kuzubas2014}
Kuzubas, T.U., Omercikoglu, I., Saltoglu, B.: Network centrality measures and
  systemic risk: An application to the turkish financial crisis. Physica A
  \textbf{405},  203--215 (2014)

\bibitem{kwapien2017}
Kwapien, J., Oswiecimka, P., Forczek, M., Drozdz, S.: Minimum spanning tree
  filtering of correlations for varying time scales and size of fluctuations.
  Physical Review E  \textbf{95}(5),  052313 (2017)

\bibitem{leon2014}
León, C., Pérez, J.: Assessing financial market infrastructures' systemic
  importance with authority and hub centrality. Journal of Financial Market
  Infrastructures  \textbf{2}(3),  67--87 (2014)

\bibitem{mantegna1999}
Mantegna, R.N.: Hierarchical structure in financial markets. The European
  Physical Journal B-Condensed Matter and Complex Systems  \textbf{11}(1),
  193--197 (1999)

\bibitem{jaramillo2014}
Martinez-Jaramillo, S., Alexandrova-Kabadjova, B., Bravo-Benitez, B.,
  Solórzano-Margain, J.P.: An empirical study of the mexican banking
  system’s network and its implications for systemic risk. Journal of
  Economic Dynamics \& Control  \textbf{40},  242--265 (2014)

\bibitem{moler2003}
Moler, C., Van~Loan, C.: Nineteen dubious ways to compute the exponential of a
  matrix, twenty-five years later. SIAM Review  \textbf{45}(1),  3--49 (2003)

\bibitem{pastor2015}
Pastor-Satorras, R., Castellano, C., Van~Mieghem, P., Vespignani, A.: Epidemic
  processes in complex networks. Review of Modern Physics  \textbf{87}(3), ~925
  (2015)

\bibitem{Peron011}
Peron, T.D., Rodrigues, F.A.: Collective behavior in financial markets. EPL
  (Europhysics Letters)  \textbf{96}(4),  48004 (2011)

\bibitem{Rodrigues019}
Rodrigues, F.A.: Network centrality: an introduction. In: A mathematical
  modeling approach from nonlinear dynamics to complex systems, pp. 177--196.
  Springer (2019)

\bibitem{rossi2018}
Rossi, A.G., Blake, D., Timmermann, A., Tonks, I., Wermers, R.: Network
  centrality and delegated investment performance. Journal of Financial
  Economics  \textbf{128},  183--206 (2018)

\bibitem{temiz2017}
Temizsoy, A., Iori, G., Montes-Rojas, G.: Network centrality and funding rates
  in the e-mid interbank market. Journal of Financial Stability  \textbf{33},
  346--365 (2017)

\bibitem{zhao2018}
Zhao, L., Wang, G., Wang, M., Bao, W.~Li, W., Stanley, H.E.: Stock market as
  temporal network. Physica A  \textbf{506},  1104--1112 (2018)

\end{thebibliography}
\end{document}